\documentclass[aip,super]{revtex4-1}
\usepackage{graphicx,amsmath}
\usepackage{amsmath,amssymb}
\usepackage[matrix,frame,arrow]{xy}
\usepackage{color}
\usepackage[vcentermath]{youngtab}
\usepackage{braket}
\usepackage{appendix}
\usepackage{siunitx}
\usepackage{mathtools}
\usepackage{url}
\usepackage{dsfont}
\usepackage{verbatim}
\usepackage{qcircuit}

\makeatletter
\newcommand{\mathleft}{\@fleqntrue\@mathmargin0pt}
\newcommand{\mathcenter}{\@fleqnfalse}
\makeatother

\newcommand{\s}{\mathfrak{s}}

\begin{document}
\date{\today}
\title{Young frames for quantum chemistry}
\author{Sahil Gulania}
\affiliation{Department of Chemistry, University of Southern California, Los Angeles, CA 90089, USA}
\author{James Daniel Whitfield} 
\affiliation{Department of Physics and Astronomy, Dartmouth College, Hanover, NH 03755, USA}

\begin{abstract}
Quantum chemistry often considers atoms and molecules with non-zero spin.  In such cases, the need for proper spin functions results in the theory of configuration state functions.  Here, we consider the construction of such wavefunctions using the symmetric group and more specifically Young projectors. We discuss the formalism and detail an example to illustrate the theory.  Additionally, we consider the pros and cons of specific implementations of spin symmetry in quantum simulation.
\begin{figure}[h!]
    \centering
    \includegraphics[scale=1.9]{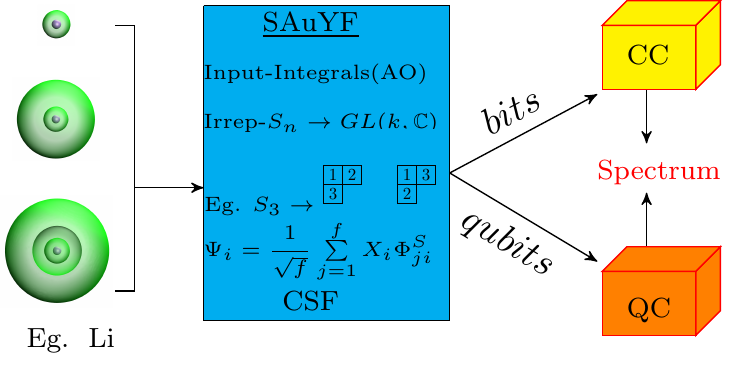}
    \label{fig:flow}
\end{figure}
\end{abstract}
\maketitle


\newpage 


Accurate models for simulating chemical system must account for inherent quantum nature.  In doing so numerically, all known methods face the challenge of intractable computation as the system grows and the accuracy requirement increases. Approximations such as the Born-Oppenheimer approximation~\cite{born1927quantentheorie,woolley1977molecular,koppel2009jahn} allow us to separate the nuclear and electronic problems.  In solving the electronic problem, further approximations are made such as the approximate universal functionals in DFT~\cite{kohn1965self} or the truncation of the Hilbert space using linear ansatz (CISD~\cite{Szabo96}, CASSCF~\cite{roos1980complete}) or using exponential ansatz (CCSD~\cite{purvis1982full}).  

The introduction of quantum computation for quantum simulation \cite{Feynman82,Lloyd96} offers the possibility of circumventing costly numerical methods.  Instead, the dynamics of one quantum system (e.g. a quantum computer) can be used to simulate another (e.g. a molecule of commercial significance). This route to quantum simulation has been one of the driving motivations for building quantum computers. Many studies and algorithms have been proposed to close the gap between quantum computers and electronic structure calculations~\cite{ward2009preparation,sugisaki2016quantum,whaley2014quantum}. This article aims to take advantage of conventional algorithms developed around the symmetric group 
and use those techniques to do the same on the quantum computer.

Given a system of $N_{nuc}$ nuclei and $N$ electrons, the non-relativistic time-independent electronic Hamiltonian~\cite{Szabo96}  can be written as, 
\begin{equation}
\mathcal{\hat{H}}_{el} = - \sum_{i=1}^{N} \dfrac{1}{2}\nabla_{i}^{2} - \sum_{i=1}^{N} \sum_{n=1}^{N_{nuc}} \dfrac{Z_{n}}{|r_{i}-R_n|} + \sum_{i=1}^{N}\sum_{j  > i}^{N}\dfrac{1}{|r_{i}-r_j|}
\label{eq:hamiltonian}
\end{equation}
where, first term represents kinetic energy of electron $i$, $Z_{m}$ is the atomic number of nucleus $m$, $r_{iI}$ represents the distance between electron $i$ and nucleus $I$, $r_{ij}$ is the distance between electrons $i$ and $j$. 
Here and throughout we are using atomic units where the numerical value of $\hbar$, $e$, $m_e$, $(4\pi \epsilon_0)^{-1}$ are set to unity.

This Hamiltonian gives the full spectrum of electronic energies at a given nuclear geometry (Born-Oppenheimer approximation).  To get potential energy curves one should also add in the fixed nuclear potential energy $\sum_{n > m}{Z_{n}Z_{m}/R_{nm}}$.
Given that the Hamiltonian is non-relativistic and the system is subject to no magnetic fields, it has no terms which explicitly depends on spin variables. Therefore Hamiltonian commutes with the total spin operator $[\mathcal{\hat{H}}_{el},\hat{S}^{2}]  = 0$ and the projection of the spin onto an axis $ [\mathcal{\hat{H}}_{el},\hat{S}_{z}] = 0$.
The commutation relations of the operators ensure a common set of simultaneous eigenfunctions for $\hat S^2$, $\hat S_z$, and $ \mathcal{\hat{H}}_{el}$.  The energy eigenstates with definite spin eigenvalues 
are be classified as singlets ($S=0$), doublets ($S=1/2$),
triplets ($S=1$) and so on.  Solutions to Hamiltonian which are also eigenfunctions of $\hat S^{2}$ with eigenvalue $S(S+1)$ will have $2S+1$ degeneracy corresponding to the eigenvalues of $\hat S_z$ when no magnetic field is present. 
As the solution to $\mathcal{\hat{H}}_{el}$ is computationally costly~\cite{pople1999nobel}, 
working in spin eigenbasis is a way to reduce the computation cost by working in subspace, which has been extensively used by physicists and chemists in standard computation~\cite{friis1996spin,mcweeny1988spin,sarma1977programmable}. Similarly, working in spin eigenbasis provides speed up for simulating electronic structure on quantum computers~\cite{Whitfield13,sugisaki2016quantum}.




The outline of the paper is as follows: first in section \ref{sec:intro} the symmetric group and its attendant ideas necessary for the construction of spin eigenfunction are detailed.  Then in section \ref{sec:example} an application to doublet Li is detailed followed by a discussion of our conclusions in section \ref{sec:conclusion}. 

In this article, we detail the construction and implementation of the spin adapted formalism including notes on our Fortran implementation. In the first section, we describe the symmetric group approach.  Then we apply the formalism to the Lithium atom. Finally, we discuss possible applications to quantum computing and future work. Readers may find the notation guide in Appendix A useful when perusing the article.

\section{Symmetric group}\label{sec:intro}

For consistency with the notion of identical particles, the observable quantities such as the probability of occupancy should be unaffected by the exchange of two particles (since they are identical).  The density matrix which contains a completed description of the system is quadratic in the wavefunction $\hat \rho=\sum p_i \ket{\psi_i}\bra{\psi_i}$.  Hence, the exchange of two particles cannot change the wavefunctions however a phase factor may be accumulated (since both the wavefunction and its dual appear).  If the phase factor resulting from the exchange of two particles is +1, then the wavefunction is unchanged and said to be symmetric under particle exchange. If instead, as in electronic wavefunction, the phase factor associated with particle exchange is $-1$, then the wavefunction is antisymmetric under particle exchange.  In both cases, the properties of this phase factor can be understood using group theory.  This section focuses on the application of the symmetric group to fermionic (antisymmetric) wavefunctions since these are most relevant to quantum chemistry.


This section is broken up into seven subsections.  First, in order to explain applications of the symmetric group to quantum wavefunctions, we begin in subsection \ref{ssec:permutations} by defining the action of permutations. Next, Young frames and their correspondence to irreducible representations of the symmetric group are introduced in subsections \ref{ssec:yframe} and \ref{sssec:rep_ortho}.  In subsection \ref{sssec:rep_conj}, the conjugate of an irreducible representations is defined as it will be needed to construct spin eigenfunctions using the symmetric group.  The spin eigenfunctions generated using the symmetric group approach (subsection \ref{sssec:spin_permute}) must be paired with possible spatial functions labelled by Weyl tableaux introduced in subsection \ref{sssec:weyl}.  Finally, in the last subsection \ref{sssec:combination}, the full antisymmetric wavefunction over both spin and spatial degrees of freedom is constructed.

%
%
%


\subsection{Permutations}\label{ssec:permutations}

Here we introduce the key notions of permutations and establish our notations.  The central object of this section is the symmetric group which must satisfy the requirements of an abstract group.  A group is a set and an associated binary operation. The group must be closed under this operation. There must be an identity element, inverses exist within the set for all group members. The final requirement is that the binary operation is associative: $A*(B*C)=(A*B)*C$.  Here and throughout, we will assume our binary operation to be a multiplicative composition as defined below.

The symmetric group of order $N$, written $S_N$, is the set of all $N!$ permutations of $N$ items.  The action of a permutation $P\in S_N$ is defined as a transformation on which 1 is replaced by $p_{1}$, 2 is replaced by $p_{2}$, ... and $N$ is replaced by $p_{N}$. We can represent this permutation in a number of ways. 
We will be using cyclic notation to specify our permutations as $(ij...k)(pq...r)(ab..c)...$.  Each parenthesis, $(ijk...l)$, represent the transformation that puts item $i$ in the $j$th location, $j$ to $k$th location, and so on with $l$ being placed in the $i$th location.  When specifying a permutation that leaves $N-m$ items unchanged, then we do not write down the trivial cycles since $N$ may be considered fixed throughout the present discussion.




As discussed earlier, the exchange of two particles will play a key role in the physical properties of that wavefunction.  The exchange of two particles corresponds to a transposition permutation.
A transposition is defined as a permutation of only two items from the possible $N$ items.
\begin{equation}
\begin{pmatrix}
    i & j  
\end{pmatrix}
=
\begin{pmatrix}
    i & j\\
    j & i  
\end{pmatrix}
\end{equation}
where $i,j \in \{1,2,...,N\}$. If $j=i+1$, then this type of transposition is known as an elementary transposition. It can be shown that any $P \in S_{N}$ can be written as product of transpositions as well as elementary transpositions.

The number of transpositions required to represent a permutation depends on the choice of transpositions used in the decomposition.  However, the even/odd parity is unique and will be denoted as $sgn(P)$.  If the number of transpositions needed in any decomposition of permutation $P$ is odd then $sgn(P)=1$ and if the transposition number is even then $sgn(P)=0$. 


Additionally, we can represent the permutations using matrices.  A collection of matrices that multiply according to the rules of the group are called a representation of that group.  The trivial representation where every permutation is represented by the identity matrix, trivially obeys the multiplication of group elements for every group. The first example of a representation of a permutation $P$ that maps item $i$ to the $p_i$th location is: 
\begin{equation}
\mathcal{M}(P)_{ij} = 
\begin{cases}
  1 & \text{if $i= p_{j} $ } \\
  0 & \text{otherwise}
\end{cases}
\end{equation}
whose action can be defined using matrix multiplication rules when considering action on items listed in a column.  For example, the matrix representation of expression $(123) \cdot abc= cab$ is
\begin{equation}
    \begin{pmatrix}
    0 & 0 & 1\\1&0&0\\0&1&0
    \end{pmatrix}
    \begin{pmatrix}
    a\\b\\c
    \end{pmatrix}
    =
    \begin{pmatrix}
    c\\a\\b
    \end{pmatrix}
\end{equation}
where $a,b,c$ are items of any sort.  The vector space corresponding to the different ordering of the three items gives rise to the representation $\mathcal{M}(P)$ for $P\in S_3$.  

To motivate the idea of irreducible subspaces, consider the equal sum of the vectors 
\begin{equation}
\ket{v}=    \begin{pmatrix}
    a\\b\\c
    \end{pmatrix}+   
    \begin{pmatrix}
    a\\c\\b
    \end{pmatrix}
    +    \begin{pmatrix}
    b\\a\\c
    \end{pmatrix}
    +    \begin{pmatrix}
    b\\c\\a
    \end{pmatrix}
    +    \begin{pmatrix}
    c\\b\\a
    \end{pmatrix}+
        \begin{pmatrix}
    c\\a\\b
    \end{pmatrix}
    \label{eq:v}
\end{equation}
Now if we consider the action of any permutation of this vector $v$, one readily notes that nothing happens since all permutations merely change the order of the summation.  This is the simplest example of an invariant subspace under the group $S_3$.  

Once one finds the subspace that is invariant, then one can remove that subspace from the original vector space resulting in a reduced space that is also invariant under the group $S_N$.  This resulting reduced space may also contain additional invariant subspaces but if it does not then it is said to be irreducible.  The space $\{ v \}$ is necessarily irreducible since it is one-dimensional.

Given any vectors space with a defined action of the permutation group then matrix representation can be found as we will do in section \ref{sssec:rep_ortho} below.





\subsection{Young frames and tableaux}\label{ssec:yframe}

The Young frame is a way to label the irreducible subspaces of the symmetric group.  Note that the Young frames also mark the irreducible subspaces of the unitary group (a consequence of the Schur-Weyl duality). In sub-section \ref{sssec:weyl} when considering the spatial functions corresponding to a particular spin state this feature will play a crucial role.  The Young frames also provide a system for labeling the basis vectors of the irreducible subspaces using an extension known as Young tableaux (\emph{singular}: tableau).  We define both followed by examples for $N=5$.
Given a number $N$, a partition~\cite{Pauncz95} is defined by $\lambda_{1}$, $\lambda_{2}$, ..., $\lambda_{k}$ such that 
\begin{equation}
\sum_{i=1}^{k}\lambda_{i} = N
\end{equation} 
and  
\begin{equation}
\lambda_{1} \geq \lambda_{2} ... \geq \lambda_{k}
\end{equation}
where $\lambda_{i}\in\{1,2,...N\}$. Diagrammatically, these partition represents frame like structures where the $k^{th}$ row has $\lambda_{k}$ boxes. They are also known as Young frames~\cite{young1934quantitative}. When numbers are filled from 1 to $N$, they are called Young tableau. For $N=5$, the possible Young frames are
\begin{align}
    \yng(1,1,1,1,1) && \yng(2,1,1,1) && \yng(2,2,1)\\
    [1 1 1 1 1] && [2 1 1 1] && [2 2 1]\nonumber
\end{align}
\begin{align}
    \yng(3,1,1) && \yng(3,2) && \yng(4,1)  && \yng(5) \\
    [3 1 1] && [3 2] && [4 1] &&  [5]\nonumber
\end{align}
When numbers are filled in increasing order within each row and column from left to right and top to bottom respectively, results in
a standard Young tableau. $f([\lambda])$ is  equal to the number of standard Young tableau
For partition [3,2], $f([3,2]) = 5$ and the standard Young tableaux are, 
\begin{align}
\young(123,45) & &     \young(124,35) && \young(134,25) &&\young(125,34) && \young(135,24) 
\label{eq:young32}
\end{align}

Each standard Young tableaux labels a basis vector of the irreducible subspace. Given a standard tableau $T_A^{[\lambda]}$, we can write down the projection operators, $E^{[\lambda]}_{AA}=\mathcal{S}_{rows(T_A)}\mathcal{A}_{cols(T_A)}$ and transfer operators $E_{AB}^{[\lambda]}=E_{AA}^{[\lambda]}P_{T_A\leftarrow T_B}$. Operators $\mathcal{S}$ and $\mathcal{A}$ are symmetrizer and antisymmetrizer operators. 
\begin{eqnarray}
    \mathcal{A}&=&\sum_{P} (-1)^{sgn(P)} P\\
    \mathcal{S}&=&\sum_{P}  P
\end{eqnarray}
Then given a list of items, $F$, that can be acted on by permutations, the vector space $\{F_A^{\lambda}=E_{AB}^{[\lambda]}\cdot F\}$ forms a basis for the irreducible subspace spanned labelled by $[\lambda]$.  
Given this as a basis, one can then obtain the irreducible representation of $P$ by examining how it transforms the basis vectors among themselves.  This circuitous route can be shortcut using Young's algorithm (see next subsection) to write down the irreducible representations directly. 

Before closing this subsection, let us consider the consequences of using the $E^{[\lambda]}$ operators to construct the irreducible subspace.  In the absence of magnetic field, the electrons have only two possible spin states we label as $\alpha$ and $\beta$.  Because the operators $E^{[\lambda]}_{AA}$ antisymmetrize the columns of tableau $T_A$, the mathematical restriction on possible irreducible subspaces limits tableaux for spin states to only two rows.  If the tableaux have three rows then a tableau of the form 
\begin{equation}
\yng(1,1,1)    
\end{equation}
Then no matter which spin function, $\alpha$ or $\beta$, is chosen, the antisymmetrization over the column results in zero.  Consequently, the first and second rows must correspond to different spin variables. The $S_z$ value of the total spin is related to the difference in length of the two rows.  Hence, for $N$ electrons with total spin $S$, the only partition of interest is $[pq] : p+q=N$ with $p-q=2S$.

\subsection{Irreducible representation of the symmetric group using Young tableaux}\label{sssec:rep_ortho}

Given a partition, the corresponding irreducible representations of the group elements can be generated using the basis vectors corresponding to standard Young tableaux. The algorithm for doing this is known as Young's algorithm.  First, we will introduce some additional terminology, then describe the algorithm and end with an example of the representation of $S_5$ in the subspace corresponding to partition [32].

The axial distance rule in a standard tableau can be used to construct the elementary transpositions using Young's algorithm~\cite{mcweeny1992methods}. 
The axial distance ($d_{pq}^{A}$) between two numbers $p$ and $q$ in a standard Young tableau $T_A^{[\lambda]}$ is defined as number of weighted steps required to reach number $q$ from number $p$ where steps to the left or downward are counted as a positive step while those to the right and upward are weighted as negative steps.


As elementary transpositions can generate all the permutations, computing irreps of elementary transposition is sufficient for generating the full group. Irreps for a given partition $[\lambda]$ and elementary transposition $(k\;k+1)$, can be obtained using the axial distance between $k$ and $k+1$ in $T_{A}^{[\lambda]}$ via
\begin{equation}
U^{[\lambda]} (k\;k+1)_{AA} = -1/d_{k\;k+1}^{A}
\end{equation}
For non-diagonal elements, apply permutation $(k \; k+1)$ to the tableau $T_{A}^{[\lambda]}$. If the result is not a standard Young tableau, then the matrix element is zero. If $(k\;k+1)T_{A}^{[\lambda]} = T_{B}^{[\lambda]}$, then
\begin{equation}
U(k\;k+1)_{AB}^{[\lambda]} = \sqrt{1-\left[U^{[\lambda]}(k\;k+1)_{AA}\right]^{2}}
\end{equation} 

In general, if the irreducible vector spaces has basis states $\{F_A^{\lambda}\}$,
then the action of a group element $P\in S_N$ must only rotate the states within
the irrep.  The irreducible matrix representation of the permutation operator $P$, 
$U^{[\lambda]}(P)$, records how the basis functions are changed within the 
irreducible subspace.  We use the following definition for the transformation of the
basis vectors within the irreducible subspace:
\begin{equation}
    P\cdot \ket{F_A^\lambda}=\sum_B^{f([\lambda])} U^{[\lambda]}(P)_{BA} \ket{F_B^\lambda}
    \label{eq:irrep_transform}
\end{equation}

For partition [32] in $S_{5}$ irreps for elementary transpositions are computed as:
\begin{align*}
U^{[32]}(1\;2)&=
\begin{pmatrix}
1\; & 0 & 0 & 0 & 0 \\
0 & 1 & 0 & 0 & 0 \\
0 & 0 & -1 & 0 & 0 \\
0 & 0 & 0 & 1 & 0 \\
0 & 0 & 0 & 0 & -1 \\
\end{pmatrix}
\hspace{2cm}
&
U^{[32]}(2\;3)&=
\begin{pmatrix}
1 & 0 & 0 & 0 & 0 \\
0 & -\frac{1}{2} & \frac{\sqrt{3}}{2} & 0 & 0 \\
0 & \frac{\sqrt{3}}{2} & \frac{1}{2} & 0 & 0 \\
0 & 0 & 0 & -\frac{1}{2} & \frac{\sqrt{3}}{2} \\
0 & 0 & 0 & \frac{\sqrt{3}}{2} & \frac{1}{2} \\
\end{pmatrix}\\[1cm]
U^{[32]}(3\;4)&=
\begin{pmatrix}
-\frac{1}{3} & \frac{\sqrt{8}}{3} & 0 & 0 & 0  \\
\frac{\sqrt{8}}{3} & \frac{1}{3} & 0 & 0 & 0 \\
0 & 0 & 1 & 0 & 0 \\
0 & 0 & 0 & 1 & 0 \\
0 & 0 & 0 & 0 & -1 \\
\end{pmatrix}
&
\hspace{2cm}
U^{[32]}(4\;5)&=
\begin{pmatrix}
1 & 0 & 0 & 0 & 0 \\
0 & -\frac{1}{2} & 0 & \frac{\sqrt{3}}{2} & 0 \\
0 & 0 & -\frac{1}{2} & 0 & \frac{\sqrt{3}}{2} \\
0 & \frac{\sqrt{3}}{2} & 0 & \frac{1}{2} & 0 \\
0 & 0 & \frac{\sqrt{3}}{2} & 0 & \frac{1}{2} \\
\end{pmatrix}
\end{align*}

Since the irreducible representations of the group allow us to construct symmetry adapted states, the matrix representations are given in this section will be of key importance in section (\ref{sssec:spin_permute}) when constructing spin eigenfunction.  

\subsection{Conjugate Representation}\label{sssec:rep_conj}
In general, cannot use the same irrep of the symmetric group for both the spatial and spin functions to obtain a totally antisymmetric wavefunction over all variables.  Even in the situation where both space and spin symmetries belong to the same irrep, their basis function must be correctly matched to achieve antisymmetry. The combined wavefunction should for any odd permutation yield a $-1$ phase factor following the anti-symmetry rule.
Hence, there is a need to construct the corresponding spatial and spins states to make sure that odd permutations of the combined coordinates under particle exchange result in $-1$ phase of total wavefunction. This can be achieved by using the representations which are conjugate to irreducible representation. 

Conjugate representation, $V^{[\tilde\lambda]}$, of any representation, $U^{[\lambda]}$, is defined such that
\begin{equation}
    U^{[\lambda]}(P) V^{[\tilde\lambda]}(P) = V^{[\tilde\lambda]}(P) U^{[\lambda]}(P) = (-1)^{sgn(P)}\mathds{1}
\end{equation} 
for all $P\in S_N$. As a result, we can also define the conjugate representation as
\begin{equation}
V^{[\tilde\lambda]}(P)_{AB} = (-1)^{sgn(P)} U^{[\lambda]}(P)_{BA}=(-1)^{sgn(P)}U^{[\lambda]}(P^{-1})_{AB}
\label{eq:V}
\end{equation}
Since we are considering orthogonal representation (both real and Hermitian), the transpose of $U^{[\lambda]}(P)^T=U^{[\lambda]}(P^{-1})$.

The conjugate irreducible space of $[\lambda]$ will be written as $[\tilde\lambda]$ where the number of rows (columns) of $\lambda$ is the number of columns (rows) of $\tilde\lambda$.  The basis vectors of the two spaces are related by taking the tableau transpose of $T_A^{[\lambda]}$ to get $T_A^{[\tilde\lambda]}$.  For instance,
\begin{align}
    \young(123,45) && \young(14,25,3)
\end{align}
Using the correspondence of the basis vectors, along with the rules from section \ref{sssec:rep_ortho}, we would also arrive at \eqref{eq:V}.  Below, in section \ref{sssec:combination}, we provide an alternative motivation for defining the conjugate representation.

With the introduction of irreducible representations of $S_N$ and their conjugate representations, we are in a position to combine the two symmetries to arrive at an antisymmetric wavefunction over space and spin. If we enforce the permutation symmetry corresponding to total spin $S$, then the conjugate irreducible subspace must then be used for the spatial components of the wavefunctions \ref{sssec:weyl}.  Before combining the spatial and spin wavefunctions, we turn to the generation of spin eigenfunction possessing the correct symmetry in the next subsection.



\subsection{Spin eigenfunctions generated with  symmetric group irreducible subspace projections}\label{sssec:spin_permute}

The electron as a structure-less point particle still has angular momentum associated with its spin.  Since there is no preferential direction for quantization, we can assume all the electrons are either aligned or anti-aligned along the $z$ axis. Then constructing eigenfunctions of $\hat S_z=\sum_{i=1}^N \hat{s}_z(i)$ is a matter of selecting how many spins are upward and how many are downward.  However, the construction of spin eigenfunctions that are also eigenfunctions of the $\hat S ^2$ operator is less straightforward. The motivation for introducing the machinery of the symmetric group and its group structure is to simplify the understanding of spin variables in electronic wavefunctions.  In this subsection, we realize this goal using Wigner projection operators. We also use branching diagrams to classify and count the spin states that correspond to a particular number of electrons and desired total spin.  First, let us recall the basics of spin eigenfunctions.

We label spin eigenfunctions for $N$ electrons, total spin $S$ and total spin projection on $z$ axis $M_{S}$ as $\Theta[S,M_S]$. Since it is a spin eigenfunction it must satisfy
\begin{equation}
    \begin{split}
        \hat{S}^{2}\Theta[S,M_S] = & S(S+1)\Theta[S,M_S]\\
        \hat{S}_{z}\Theta[S,M_S] = & M_{S}\Theta[S,M_S], \hspace{1cm} -S\leq M_{S} \leq S
        \end{split}
\end{equation}
The single-electron spin operators (lowercase) and the total spin operators are related by 
$\hat S_z=\sum_i^N \hat s_z(i)$ and $\hat S^2=\vec S.\vec S=\sum_{i,j}^N \vec s(i)\cdot \vec s(j)$.  

Let $\alpha(\s)$ and $\beta(\s)$ be one-electron functions of the spin coordinate $\s\in\{\uparrow,\downarrow\}=\{m_s=1/2,m_s=-1/2\}$ which we choose as spin eigenfunctions of $\hat s_z$.  
We let them act as indicator functions with $\alpha(1/2)=\beta(-1/2)=1$ and $\alpha(-1/2)=\beta(1/2)=0$. 
We label arbitrary one-electron spin eigenfunction ($S=1/2$) with $\sigma(\s_i)$ as 
\begin{equation}
\sigma(\s_i)=
\begin{cases}
\alpha(\s_i)\\
\beta(\s_i)
\end{cases}
\end{equation}
here $i$ refers to the $i$th electronic spin coordinate $\s_i\in\{\uparrow,\downarrow\}$.



The total number linearly independent spin eigenfunctions which gives same eigenvalue for 
$\hat{S}^{2}$ for a given configuration of $N$ is known as the degeneracy of spin eigenfunction denoted by 
\begin{equation}
f(N,S)={N\choose \frac N2-S} - {N\choose \frac N2 - S -1}
\end{equation}
Unless, magnetic field acts on the molecule, the energy spectrum remain degenerate with the subspace of the $f(N,S)$ spin eigenfunctions.

Primitive spin functions (PSF) for $N$ electrons are defined as product of one electron spin eigenfunction. 
\begin{equation}
\theta(\s_1,\s_2,...\s_N) = \sigma(\s_1)\sigma(\s_2)...\sigma(\s_N)
\end{equation}
If there are $p$ orbitals with $\alpha$ spin and $q$ orbitals with $\beta$ spin in $\theta_{k}$, then $\hat{S}_{z} \theta$ = $\frac{1}{2}(p-q)\theta$.

Graphical representation can be given to these PSF by assigning the direction of $45\si{\degree}$ to $\alpha$ and $-45\si{\degree}$ to $\beta$. These diagrams are also known as path diagrams. Also, a digital representation can be computed by assigning 1 to $\alpha$ and 2 to $\beta$. Spin functions ending in $\beta$ are ordered before the functions ending in $\alpha$.  
If the last digit is the same, then one compares the second to last and continues similarly until all functions are ordered. This is known as the last-letter sequencing, and other orderings are possible e.g., dictionary sequencing.  
See table \ref{tbl:table1} for an example of the last-letter sequencing of primitives for the $N=5$ $M_S=\frac12$ case.

\begin{table}[h!]
\centering
\begin{tabular}{|c|c|c|}
\hline
No. & $\theta(\s_1,...,\s_5)$ & Ordering \\ \hline
1 & $\alpha\alpha\alpha\beta\beta$ & 11122 \\ \hline
2 & $\alpha\alpha\beta\alpha\beta$ & 11212 \\ \hline
3 & $\alpha\beta\alpha\alpha\beta$ & 12112 \\ \hline
4 & $\alpha\alpha\beta\beta\alpha$ & 11221 \\ \hline
5 & $\alpha\beta\alpha\beta\alpha$ & 12121 \\ \hline
\end{tabular}
\caption{Last letter sequencing of the primitive spin functions of $N=5$, $M_S=\frac12$.}\label{tbl:table1}
\end{table}

\begin{figure}[h!]
    \centering
    \includegraphics[scale = 0.4]{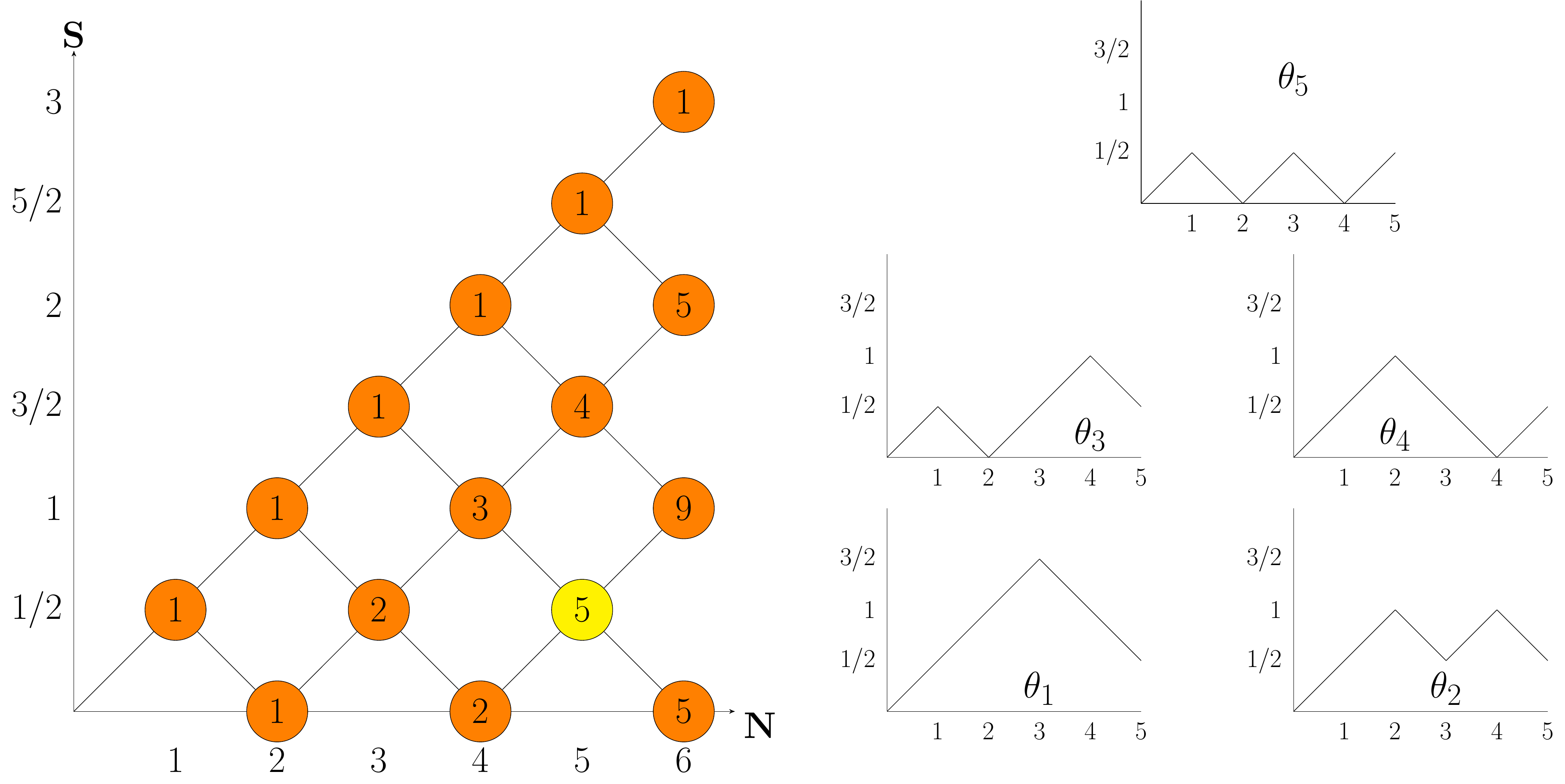}
    \caption{Branching diagram and path diagrams for $N$ electrons with total spin $S$.  The point $(N=5,S=\frac{1}{2})$ is highlighted and corresponds to the five path diagrams on the left.  The five path diagrams correspond to the PSF shown in Table~\ref{tbl:table1}.}
    \label{fig:branch}
\end{figure}

The number of PSFs having whole path diagram above the x-axis gives the degeneracy of spin configuration,  $f(N,S)$, defined in the previous subsection. It is also equal to the dimension of the irreducible representation generated by the Young frame of partition $[pq]: p+q=N; p-q=2S$. Degeneracy of spin eigenfunction can be represented using a branching diagram where each point is assigned a number equal to total PSF which are above the x-axis. For $N=5$ and $M_{S}=1/2$ there can be eight PSF, but only five of them will be above the x-axis. They are shown in Fig. \ref{fig:branch}. 

Using the primitive spin functions which stay above the x-axis and irreducible representation of $S_N$ within 
partition $[pq]$, spin eigenfunctions can be constructed using the Wigner projection operator~\cite{Pauncz95}.  The projectors onto the irreducible subspace are connected to the irreducible representation of the permutation group via
\begin{eqnarray}
\mathbf{P}^{[\lambda]AB}=\sum_{g\in G} U^{[\lambda]}(g^{-1})_{AB}\;   g=\sum_{g\in G} U^{[\lambda]}(g)_{BA}\;   g
\label{eq:wigner-projector}
\end{eqnarray}
Here $U^{[\lambda]}(g)_{AB}$ is the $A,B$ entry of the matrix representation of group element $g$ within the irreducible subspace.  For example, in the trivial representation discussed after \eqref{eq:v} $U^{[111]}(P)=1$ for all $P\in S_N$.  Then the Wigner projector is
\begin{equation}
    \mathbf{P}^{[111],A=1,B=1}=\sum_{P\in S_N} P\propto \mathcal{S}
\end{equation}
where $\mathcal{S}$ is the symmetrizer as anticipated by previous experience.

As a consequence of the orthogonality theorems of group theory, the Wigner projection operators transform as basis function of the irreducible subspace (see the appendix for an elementary introduction).  Consequently, we may write the projection operator as 
\begin{equation}
\mathbf{P}^{[\lambda] AB }=\ket{F^{\lambda}_B}\bra{F^{\lambda}_A}
\label{eq:projector}
\end{equation}
where $F_I^\lambda$ represents the $I$th basis function in the carrier vector space of irrep $[\lambda]$.  

Then with the irreducible representations known, we will obtain the following rules 
for transformation of the Wigner projection operators:
\begin{eqnarray}
P \cdot \mathbf{P}^{[\lambda] AB}
&=&P\cdot  \ket{F_B^\lambda}\bra{F_A^\lambda} 
=\sum_I U^{[\lambda]} (P)_{IB}\ket{F_I^\lambda}\bra{F_A^\lambda}\\
&=&\sum_I U^{[\lambda]} (P)_{IB} \mathbf{P}^{[\lambda] AI}
\end{eqnarray}
And
\begin{eqnarray}
 \mathbf{P}^{[\lambda]AB}\cdot P
&=& \ket{F_B^\lambda}\bra{F_A^\lambda} \cdot P 
 = \ket{F_B^\lambda}\left(P^{-1} \cdot \ket{F_A^\lambda} \right)^\dag\\
&=&\ket{F_B^\lambda} 
      \left(\sum_I U^{[\lambda]}(P^{-1})_{IA}\ket{F_I^\lambda}\right)^\dag\\
&=&\sum_I U^{[\lambda]}(P^{-1})_{IA}\ket{F_B^\lambda}\bra{F_I^\lambda}\\
\mathbf{P}^{[\lambda]AB}\cdot P&=&\sum_I U^{[\lambda]} (P^{-1})_{IA} \mathbf{P}^{[\lambda] IB}\label{eq:31}
\end{eqnarray}
We have assumed that $U^{[\lambda]}$ is real and orthogonal.  This is justified 
since we are using Young's orthogonal representation above.
Specializing to the case of the spin eigenfunctions, we can get the projection onto a basis function of the irreducible space. For an arbitrary primitive $\theta(\s_1,...\s_N)$ we have
\begin{equation}
    \Theta_{A}(N,S,M_S) =  N_{C}\left[\sum_{P} U^{[\lambda]}(P^{-1})_{CA}\; P\cdot \theta\right]=N_{C}\;\mathbf{P}^{[\lambda]CA}\cdot \theta
    \label{eq:construct}
\end{equation}
where $N_C$ is a normalization constant which depends on the fixed value of $C$.  Each fixed value of $C$ generates the same projection, see \eqref{eq:projector}.  We have also used fact that $U(P^{-1})_{AB}=U(P)_{BA}$ for orthogonal representations. Conversely, if the spin eigenfunctions corresponding to the irrep $[\lambda]$ are known, they can generate the irreducible representations. One constructs the representations $U^{[\lambda]}(P)$ from the transformation rules of irreducible representations 
\begin{equation}
    P\cdot \Theta_{A} = \sum_{B=1}^{f([\lambda])}\Theta_{B}\; U^{[\lambda]}(P)_{BA}
\end{equation}

\subsection{Weyl tableaux}\label{sssec:weyl}
One can construct $2m \choose N$ possible $N$-electron product states using $m$ possible one-electron spatial functions.  
Spin adaptation reduces the computation by working with eigenfunctions of $\hat{S}^{2}$.  However, there is still a need to make the selection within the $M$ one-electron function and combine them in spin adapted wavefunctions. 

The entries in Weyl tableau are from 1 to $M$, where $M$ is the total number of basis functions.  
Entries in each row may be equal or increase from left to right.  In each column, entries must strictly increase downward.  For a given $N, S, M$, the number of Weyl tableaux is given by
\begin{equation}
    W(N,S,M) = \dfrac{2S+1}{M+1}\binom{M+1}{N/2+S+1}\binom{M+1}{N/2-S}
\end{equation}
For $N=5,S=1/2,M=7$ there can be 490 Weyl tableau or CSF
\begin{equation}
    \young(ab,cd,e)  ,\quad \begin{aligned} a&\le b,\\ c&\le d,\\a&<c<e, \\ b&<d    \end{aligned}
\end{equation}
where $a,b,c,d,e\in\{1,2,...,7\}$.

The Weyl tableaux for $N=3$, $S=1/2$ will belong to partition [21] 
\begin{equation}
    \young(ij,k) ,\quad \begin{aligned} j&\ge i,\\ k&> i\end{aligned}
\end{equation}
If three basis functions are used then $W(3,1/2,3) = 8$. All possible Weyl tableaux corresponding to CSFs are shown below
\begin{align}
\young(11,2) & &     \young(11,3) && \young(12,2) &&\young(13,3) && \young(22,3) && \young(23,3) && \young(12,3) && \young(13,2)
\end{align}

\subsection{Combination of Spatial and Spin function}\label{sssec:combination}
After constructing the set of orthonormal spin eigenfunctions, $\{\Theta_{A}\}$ per 
\eqref{eq:construct}, the corresponding spatial part of the wavefunction must be 
constructed with the correct symmetry.  Similar to the previous section, the 
primitive spatial wave function is product function of one-electron spatial orbitals
\begin{equation}
    Y[abc...](r_1,r_2,r_3,...) = \phi_a(r_{1})\phi_b(r_{2})\phi_c(r_{3})... 
\end{equation}

We can find the spatial wavefunction with the correct symmetries by performing total antisymmetrization of a spatial primitive $Y$ and an arbitrary spin eigenfunction $\Theta_D$. The unnormalized wavefunction is then
\begin{eqnarray}
\Psi(r_1,\sigma_1,r_2,\sigma_2,...)&=&  \mathcal{A} (Y \Theta_{D})\\
&=&\sum_{P} (-1)^{sgn(P)} (P^{r}\cdot  Y) (P^{\s} \cdot \Theta_D)
\end{eqnarray}
Permutation $P$ acts on both the spin and spatial components of the wavefunction with  $P^{\s}$ acting as permutation $P$ on the spin component and $P^{r}$ acting on the spatial parts. 

From eq. \eqref{eq:irrep_transform} in section \ref{sssec:spin_permute}, the action of a permutation on spin eigenfunction that carries the representation is
\begin{equation}
P^{\s} \cdot \Theta_{D} = \sum_{A=1}^{f([\lambda])} \Theta_{A} U^{[\lambda]}(P)_{AD}
\label{eq:comb2}
\end{equation}
Using this in the previous equation gives
\begin{eqnarray}
\Psi &=&\sum_{P} (-1)^{sgn(P)} (P^{r}\cdot  Y) \left(\sum_{A} \Theta_{A} U^{[\lambda]}(P)_{AD}\right)\\
&=&\sum_A \left( \sum_{P} \{(-1)^{sgn(P)}U^{[\lambda]}(P)_{AD}\} P^{r}\right)\cdot  Y  \Theta_{A} \\
&=&\sum_A \sum_P \{V^{[\tilde\lambda]}(P^{-1})_{AD}\}P^r\cdot Y \; \Theta_{A}\\
&=&\sum_A \left(\mathbf{P}^{[\tilde\lambda]DA}\cdot Y\right) \left( \mathbf{P}^{[\lambda]CA}\cdot \theta\right) \label{eq:combine_ss}
\end{eqnarray}
Here both $C$ and $D$ are arbitrary basis labels from the irreducible vector spaces $[\lambda]$ and $[\tilde\lambda]$ respectively.

\section{Examples: Li (S=1/2)}\label{sec:example}
In this section, the doublet ($S=1/2$) spin-adapted wavefunction for the Li atom is constructed using the techniques describe in the previous section. We can use the results from the end of \ref{ssec:yframe}, $N=3=p+q$ and $p-q=2S=1$, to see that partition $[pq]=[21]$ is the relevant irreducible subspace of $S_{3}$. The irreps generated using Young's algorithm are used to construct spin eigenfunctions, and the corresponding conjugate irreps are used to build spin-adapted spatial functions with the help of Weyl tableaux. In the end, the combined wavefunctions can be used to obtain the energy spectrum within a basis of three spatial Gaussian functions. 

The electronic Hamiltonian for Li atom according to eq. (\ref{eq:hamiltonian})
\begin{equation}
    \mathcal{\hat{H}}_{el}^{Li} = -\sum_{i=1}^{3}\dfrac{1}{2}\nabla_{i}^{2} - \sum_{i=1}^{3} \dfrac{3}{r_{i}} 
    + \sum_{i=1}^{3}\sum_{j>i}^{3}\dfrac{1}{r_{ij}}
\end{equation}
rewriting this as
\begin{equation}
    \mathcal{\hat{H}}_{el}^{Li} = \sum_{i=1}^{3} \hat h(r_i) + \sum_{i=1}^{3}\sum_{j>i}^{3}\hat e(r_i,r_j)
\end{equation}
where
\begin{align}
    \hat{h}(r_i) &= -\dfrac{1}{2}\nabla_{i}^{2} + \dfrac{3}{r_{i}}\\
    \hat{e}(r_i,r_j) &= \dfrac{1}{r_{ij}}
\end{align}
$\hat h(r_i)$ represents the one-electron operator and $\hat e(r_i,r_j)$ represents the two-electron operator. This notation helps in evaluating matrix elements. For a function written as product of one-electron basis functions, 
\begin{align}
    Y[abc]  &= \phi_{a}(r_1)\phi_{b}(r_2)\phi_{c}(r_3)\\
    Y[jkl] &=  \phi_{j}(r_1)\phi_{k}(r_2)\phi_{l}(r_3)
\end{align}
where $a,b,c,j,k,l$ correspond to the one-electron basis functions and $r_1,r_2,r_3$ represents electronic spatial coordinates. Expectation value of one electron and two electron operator is easy to evaluate using the following rules
\begin{align}
    \braket{Y[abc]|\hat h(r_1)|Y[jkl]} &= \braket{\phi_{a}|\hat h(r_1)|\phi_{a}}
    \braket{\phi_{b}|\phi_{k}}\braket{\phi_{c}|\phi_{l}} \\
    \braket{Y[abc]|\hat h(r_2)|Y[jkl]} &= \braket{\phi_{b}|\hat h(r_2)|\phi_{k}}
    \braket{\phi_{a}|\phi_{j}}\braket{\phi_{c}|\phi_{l}} \\
    \braket{Y[abc]|\hat h(r_3)|Y[jkl]} &= \braket{\phi_{c}|\hat h(r_3)|\phi_{l}}
    \braket{\phi_{a}|\phi_{j}}\braket{\phi_{b}|\phi_{k}}
\end{align}
similarly 
\begin{align}
    \braket{Y[abc]|\hat e(r_1,r_2)|Y[jkl]} &= \braket{\phi_{a}\phi_{b}|\hat{e}(r_1,r_2)|\phi_{j}\phi_{k}}
    \braket{\phi_{c}|\phi_{l}}\\
    \braket{Y[abc]|\hat{e}(r_1,r_3)|Y[jkl]} &= \braket{\phi_{a}\phi_{c}|\hat{e}(r_1,r_3)|\phi_{j}\phi_{l}}
    \braket{\phi_{b}|\phi_{k}}\\
    \braket{Y[abc]|\hat{e}(r_2,r_3)|Y[jkl]} &= \braket{\phi_{b}\phi_{c}|\hat{e}(r_1,r_2)|\phi_{k}\phi_{l}}
    \braket{\phi_{a}|\phi_{j}}
\end{align}
These relations can be extended to any electronic Hamiltonian. Now moving to the next step of constructing spin adapted wavefunction using standard Young tableaux with partition $[21]$. There are only two standard Young tableau, which are shown below
\begin{equation}
T_{1}= \young(12,3) \hspace{4cm} T_{2} = \young(13,2)
\end{equation}
Using the axial distance rule defined in section \ref{sssec:rep_ortho}, the irreps corresponding to elementary transpositions are
\begin{equation}
U^{[21]}(1\;2)=
\begin{pmatrix}
1 & 0  \\
0 & -1  \\
\end{pmatrix}
\hspace{2cm}
U^{[21]}(2\;3)=
\begin{pmatrix}
-\frac{1}{2} & \frac{\sqrt{3}}{2}  \\
\frac{\sqrt{3}}{2} & \frac{1}{2}  \\
\end{pmatrix}
\end{equation}
Given that $(13) = (12)(23)(12)$, $(123)=(12)(23) $ and $(132) = (13)(32)$ the remaining representations can be written as 
\begin{equation}
U^{[21]}(1\;3)=
\begin{pmatrix}
-\frac{1}{2} & -\frac{\sqrt{3}}{2}  \\
-\frac{\sqrt{3}}{2} & \frac{1}{2}  \\
\end{pmatrix}
\hspace{0.5cm}
U^{[21]}(1\;2\;3)=
\begin{pmatrix}
-\frac{1}{2} & \frac{\sqrt{3}}{2}  \\
-\frac{\sqrt{3}}{2} & -\frac{1}{2}  \\
\end{pmatrix}
\hspace{0.5cm}
U^{[21]}(1\;3\;2)=
\begin{pmatrix}
-\frac{1}{2} & -\frac{\sqrt{3}}{2}  \\
\frac{\sqrt{3}}{2} & -\frac{1}{2}  \\
\end{pmatrix}
\end{equation}
Here $U^{[21]}(i,j,k...l)$ is the representation of the permutation that takes $i$ to $j$, $j$ to $k$,..., and $l$ to $i$ (cyclic notation).  
 
\begin{table}[h]
\centering
\begin{tabular}{|c|c|r|r|c|c|c|}
\hline
$P$      & $U_{11}$  & $U_{21}$        & $U_{12}$          & $U_{22}$  & $P \cdot abc$ & $(-1)^{sgn(P)}P\cdot abc$ \\ \hline
E      & 1  & 0            & 0            & 1  & abc &abc              \\ 
(12)   & 1  & 0            & 0            & -1  & bac &-bac             \\ 
(13)   & -1/2  & $-\sqrt{3}/2$           & $-\sqrt{3}/2$          & 1/2   & cba &-cba             \\ 
(23)   & -1/2  & $\sqrt{3}/2$            & $\sqrt{3}/2$            & 1/2   & acb &-acb             \\ 
(123)  & -1/2  &$-\sqrt{3}/2$           &$\sqrt{3}/2$            & -1/2  & cab &cab              \\ 
(132)  & -1/2  & $\sqrt{3}/2$           &$-\sqrt{3}/2$            & -1/2  & bca &bca              \\ 
\hline
\end{tabular}
\caption{Coefficients of $\Theta_{A}$ and $\Phi_{J}$}
\end{table}
Using eq. (\ref{eq:construct}), spin eigenfunctions corresponding to $\lambda=[21]$ can be constructed using $\theta=\alpha\beta\alpha$ 
\begin{equation}
\begin{split}
    \Theta_{1}[\alpha\beta\alpha] &=\mathbf{P}^{[\lambda]11}\cdot\theta_1= \sum_{P}U^{[\lambda]}(P)_{11}P\cdot \theta_{1} \xrightarrow[\text{normalized}]{} \dfrac{1}{\sqrt{6}} (-2\alpha\alpha\beta +\alpha\beta\alpha + \beta \alpha \alpha) \\
    \Theta_{2}[\alpha\beta\alpha] &=\mathbf{P}^{[\lambda]12}\cdot\theta_1= \sum_{P}U^{[\lambda]}(P)_{21}P\cdot \theta_{1} \xrightarrow[\text{normalized}]{} \dfrac{1}{\sqrt{2}} (\beta\alpha\alpha-\alpha\beta\alpha)\\
    \Theta_{1}^{'}[\alpha\beta\alpha] &=\mathbf{P}^{[\lambda]21}\cdot\theta_1= \sum_{P}U^{[\lambda]}(P)_{12}P\cdot \theta_{1} \xrightarrow[\text{normalized}]{} \dfrac{1}{\sqrt{6}} ( 2\alpha\alpha\beta -\alpha\beta\alpha - \beta \alpha \alpha) \\
    \Theta_{2}^{'}[\alpha\beta\alpha] &=\mathbf{P}^{[\lambda]22}\cdot\theta_1= \sum_{P}U^{[\lambda]}(P)_{22}P\cdot \theta_{1} \xrightarrow[\text{normalized}]{} \dfrac{1}{\sqrt{2}} (\alpha\beta\alpha-\beta\alpha\alpha)
\end{split}
\end{equation}


Let $a,b,c$ denotes the one electron basis functions and using eq.(\ref{eq:combine_ss}) the total spin adapted function for $Y=abc$ can be written as,
\begin{equation}
    \Psi[abc;\alpha\beta\alpha] = \Phi_{1}[abc] \times \Theta_{1}[\alpha\beta\alpha]\quad +\quad \Phi_{2}[abc]\times \Theta_{2}[\alpha\beta\alpha] \label{eq:spin1}
\end{equation}
or
\begin{equation}
    \Psi'[abc;\alpha\beta\alpha] =  \Phi'_{1}[abc] \times \Theta'_{1}[\alpha\beta\alpha]\quad +\quad \Phi'_{2}[abc]\times \Theta'_{2}[\alpha\beta\alpha] \label{eq:spin2}
\end{equation}
where,
\begin{equation}
  \begin{split}
    \Phi_{1}[abc] &=\mathbf{P}^{[\tilde\lambda]11}\cdot Y[abc] \\
    &= \sum_{P}U^{[{\lambda}]}(P)_{11}(-1)^{sgn(P)}P\cdot Y[abc]  \xrightarrow[\text{normalized}]{} \dfrac{1}{\sqrt{12}}(2abc-2bac+cba+acb-cab-bca)\\
    \Phi_{2}[abc] &= \mathbf{P}^{[\tilde{\lambda}]12}\cdot Y[abc] \\
    &=\sum_{P}U^{[{\lambda}]}(P)_{21}(-1)^{sgn(P)}P\cdot Y[abc] \xrightarrow[\text{normalized}]{} 
    \dfrac{1}{2} (cba-acb-cab+bca)\\
    \Phi'_{1}[abc] &= \mathbf{P}^{[\tilde{\lambda}]21}\cdot Y[abc]\\
    & =\sum_{P}U^{[{\lambda}]}(P)_{12}(-1)^{sgn(P)}P^{-1}\cdot Y[abc] \xrightarrow[\text{normalized}]{} 
    \dfrac{1}{2} (cba-acb+cab-bca)\\
    \Phi'_{2}[abc] &= \mathbf{P}^{[\tilde{\lambda}]22}\cdot Y[abc]\\
    &= \sum_{P}U^{[{\lambda}]}(P)_{22}(-1)^{sgn(P)}P^{-1}\cdot Y[abc] \xrightarrow[\text{normalized}]{}
    \dfrac{1}{\sqrt{12}}  (2abc+2bac-cba-acb-cab-bca)
  \end{split}
\end{equation}
\subsection{Linearly independent set}\label{sssec:example_ortho}
Now, both the spin eigenfunction and their corresponding spin-adapted spatial are available for Li ($S=1/2$). One can use either of the equations (\ref{eq:spin1}) or (\ref{eq:spin2}) to construct the Hamiltonian matrix. We construct spin-adapted functions $\Psi$ using $ \Phi_{1}$ and $\Phi_{2}$ 
\begin{equation}
    \Psi[abc] 
    =  \Phi_{1}[abc] \times \Theta_{1} \quad 
      +\quad \Phi_{2}[abc] \times \Theta_{2}
\end{equation}

The input of primitive spatial function [$a,b,c$] to the spin-adapted function comes from Weyl tableaux which are already available in section \ref{sssec:weyl}. There are eight primitive spatial functions as shown in eq. (\ref{eq:weyl}), which will generate eight spin adapted functions. 
\begin{align}
\young(11,2)  \rightarrow  \Phi[121] && \young(11,3)  \rightarrow  \Phi[131] 
&& \young(12,2)  \rightarrow  \Phi[122] \nonumber \\
\young(13,3)  \rightarrow  \Phi[133]  && \young(22,3)  \rightarrow  \Phi[232] 
&& \young(23,3)  \rightarrow  \Phi[233] \nonumber \\
\young(13,2)  \rightarrow  \Phi[123] &&  \young(12,3)  \rightarrow  \Phi[132]
\label{eq:weyl}
\end{align}
The ordering used here is 
$\young(ac,b)  \rightarrow  \Phi[abc]$
which matches our use of 
$\young(\sigma\tau,\mu)  \rightarrow  \Theta[\sigma\tau\mu]$. Using orthonormal one-electron basis to construct three electron wavefunctions result in  non-orthogonality of  $\Psi[123]$ and $\Psi[132]$. In order to obtain the spectrum of Li (S=1/2),
one needs to orthogonalize the wavefunction and transform the Hamiltonian accordingly.
If the last wavefunction $\Psi[132]$ is constructed using eq. (\ref{eq:spin2}) as $\Psi'[123]$, then all eight wavefunctions are orthogonal and diagonalization of Hamiltonian directly provides eigen-spectrum, without any extra transformation. Instead of orthogonalizing one electron
basis and later orthogonalizing the $N$-electron wavefunction, one can stick to procedure of computing the Hamiltonian by wavefunctions constructed using eq.(\ref{eq:spin1}) in normalized basis. At the end
transforming the Hamiltonian by orthogonalizing the $N$-electron wavefunctions.

\subsection{Results}
Using three local basis functions for Li of the form 
\begin{equation}
    \phi(r)=\sum_{k=1}^3 d_k \phi^{GF}_{\alpha_k}(r)
\end{equation}
with $\phi^{GF}_{\alpha_k}(r)=(2\alpha_k/\pi)^{3/4}\exp(-\alpha_k|r|^2)$.  
The coefficients and exponents used for the basis functions are given in table \ref{table:table2}.

\begin{table}[h!]
\begin{minipage}[b]{.50\textwidth}
   \centering
\begin{tabular}{|c|c|c|}
\hline
No. & Orbital exponent & Expansion coefficients \\ \hline
1 & 36.8382 & 0.0696686 \\ 
  & 5.48172 & 0.381346  \\
  & 1.11327 & 0.681702  \\ \hline
2 & 0.540205 &-0.263127  \\
  & 0.1022550 & 1.143390  \\ \hline
3 & 0.0285650 & 1\\\hline
\end{tabular}
\caption{Local basis functions used for lithium computation.}\label{table:table2}
\end{minipage}\qquad
\begin{minipage}[b]{.380\textwidth}
   \centering
\begin{tabular}{|c|c|}
\hline
State & Energy (a.u)  \\ \hline
1 & -7.38158168 \\ 
2 & -7.18378506 \\
3 & -5.25001686 \\ 
4 & -5.02455280 \\
5 & -4.99720072 \\ 
6 & -4.71527185 \\
7 & -1.66938454 \\
8 & -1.28338664 \\\hline
\end{tabular}
\caption{Spectrum of  Li atom}\label{table:spectrum}
\end{minipage}
\end{table}

The spectrum obtained match the exact diagonalization. This can be verified using our code\cite{sauyf}.
      

%
%
%

\section{Conclusion}\label{sec:conclusion}
We expect that direct implications for the present work lie in applications to quantum computing where the exponential size of the matrix is not a problem.  Future work entails constructing spin eigenstates on quantum computers and extending the present analysis to configuration interaction black box simulation methods \cite{Toloui13,Babbush18}.

For black box quantum simulation, the sparsity can be defined with $d$ as the maximum number of non-zero elements in a row. Then the black box algorithms scale polynomially with $d$, the simulation time and the norm of the Hamiltonian.  For example, in the quantum simulation via the Taylor expansion method~\cite{Berry15} has scaling that quadratic in $d^2$ (with additionally logarithmic factors) while the qubitization approach to quantum simulation \cite{Low17} has scaling linear in $d$.  Hence any reduction in the number of matrix elements per row will translate into faster algorithms.  In the present context, the sparsity will be smaller within the projected space.  Future work will need to create an oracle function to extract single matrix elements. 
    
Additionally, we would like to consider the construction of spin eigenfunction is future work. There are two approaches to constructing spin adapted eigenstates: branching diagrams constructive approaches \cite{sugisaki2016quantum}, Serber construction \cite{Sugisaki2019} or projection methods based on phase estimation \cite{Whitfield13}.  Comparing their performance on near-term quantum computers is among the next immediate steps continuing on the present work. 

{\bf Acknowledgments}:  We thank D. Klein for helpful discussions. SG 
would like to thank USC for Dornisfe Graduate School Fellowship. JDW 
is supported by the NSF under grant number 1820747 and by the DOE, Office
of Science, Office of Advanced Scientific Computing Research, under the Quantum
Computing Application Teams program.

\bibliography{spin}

\newpage
\appendix
\section{Notation}\label{appendix:notation}
\begin{itemize}
    \item $N$ is the number of electrons, $M$ the number of basis functions.
    \item $a,b,c$ items, orbital labels $a,b,c\in\{1...M\}$. Orbital labels $a,b,c\in\{1...N\}$ when discussing $S_N$.
    \item $\phi_m(r)$ labels the $m$th spatial orbital with $r$ as a spatial location.
    \item $i,j,k$ labelling electrons. $i,j,k\in\{1,...,N\}$.
    \item $A,B,C,D$ label basis vectors of irreducible subspace (carrier space).
    \item $[\lambda]$ index for a Young frame labelling an irreducible subspace and $[\tilde \lambda]$ is the transpose of the Young frame. For example, $[\lambda]=[1224]$.
    \item $[pq]$ is the frame for $p+q=N$ and $p-q=2S$.
    \item $\s\in\{\pm1/2\}$ is the spin coordinate, $r\in \mathcal{R}^3$ is the spatial coordinate, and $x=(r,\s)$ denotes the combined coordinates.
    \item $F_k$ demotes the $k$th basis function of an arbitrary vector space.
    \item $F_K^{[\lambda]}$ denotes the function which transforms like the $K$th basis function of the irreducible subspace labelled by $[\lambda]$.
    \item $P$, a permutation. $P^r$ is a permutation of spatial coordinates, $P^{\s}$ is a permutation of spin coordinates labels.  
    \item $X(\s_1,...\s_N)$ is a primitive spin function that is a product of one-electron spin functions, $\alpha(\s)$ or $\beta(\s)$.
    \item $\Theta_{A}[S,M_S](\s_1,...\s_N)$ is $A$th symmetry adapted basis of irreducible space labelled by $[\lambda]=[N/2+S,N/2-S]$.
    \item $Y[abc..](r_1,r_2,...,r_N)$ is the primitive (product) function of $\phi_{a}(r_1)$, $\phi_{b}(r_2)$,...
    \item $\Phi_{J}[abc...;[\lambda]](r_1,r_2,...,r_N)$ is the symmetry-adapted spatial wavefunction transforming as the $J$th basis vector of irreducible subspace $[\lambda]$.
    \item $U^{[\lambda]}(g)_{AB}$ and $U^{[\lambda]}_{AB}(g)$ denote the matrix element in column $A$ and row $B$ of a representation of a group element $g$ within the irreducible subspace labelled by $[\lambda]$. 
    \item $\sigma$ is an arbitrary spin function $\alpha$ or $\beta$.
    \item $T_A^{[\lambda]}$ labels the $A$th standard Young tableau with shape $[\lambda]$. $\tilde T_J^{[\lambda]}$ labels the $J$th standard Weyl tableaux. 
    \item $sgn(P)$ gives one or zero depending on the sign of the permutation.
    \item $\Psi[abc...; S, M_S]$ is an anti-symmetric $N$-body wavefunction composed of spatial orbitals $a$, $b$, $c$ and expectation values $S$ and $M_S$ of total spin operators $\hat S^2$ and $\hat S_z$.
    \item $\mathbf{P}^{[\lambda]ij}$ is the Wigner projection operator defined by $\sum_{g\in G} U^{[\lambda]}(g^{-1})_{ij}\;  g=\frac h{D_\lambda}\ket{F_j^{[\lambda]}}\bra{F_i^{[\lambda]}}$.
    \item $\hat {A}$ indicates an $N$-body operator.
    \item $\mathcal{S}$, $\mathcal{A}$ are symmetrizer and antisymmetrizer respectively.
    \item $f([\lambda])$ is  equal to the number of standard Young tableau. 
    \item $D_\lambda$ is the dimension of the irreducible subspace.  For the symmetric group $D_\lambda=f([\lambda])$.
    \item $W_A^{[\lambda]}$ labels the $A$th  Weyl tableau with shape $[\lambda]$.
    \item $d_{k\;k+1}^{A}$ distance in the A$^{th}$ Young tableau between $k$ and $k+1$.
    \item We use $G$ for a group, $h$ for the order of the group (the number of group elements) and denote the set of irreducible representation labels as $\hat G$.
\end{itemize}

\section{SI-Branching diagram approach to spin eigenfunctions}\label{sec:branching}
If the spin eigenfunctions of an $N$ electron system are known then the branching rules can be used to construct the spin 
eigenfunctions of an $N+1$ electrons.  The branching rules can be derived, e.g. using L\"{o}wdin operator~\cite{lowdin1964angular,sugisaki2016quantum}.  The spin-eigenfunction constructed this way are also known as branching diagram eigenfunction.   

Consider spin eigenfunction $\Theta[S,M_S]$ for $N$ electrons with total spin $S$ and $z$ spin projection of $M_S$ units.
Appending the single electron spin function $\alpha$ to the spin function of $N$ electrons produces a state, whose total function is a superposition of $\Theta[S+\frac12,M_S+\frac12]$ and 
$\Theta[S-\frac12,M_S+\frac12]$. Therefore,
\begin{equation}
    \Theta[S,M_S]\otimes \alpha = c_{1}\Theta[S+\frac12,M_S+\frac12] + c_{2}\Theta[S-\frac12,M_S+\frac12]
    \label{eq:sp1}
\end{equation}

Next, we have to remove undesired spin-components of the wavefunction. We can do so by specifying an operator whose null space corresponds to the spin sector we are removing. Consider
\begin{equation}
     \mathcal{\hat O}_{s} =  {\hat S}^{2} - s(s+1)\mathbf{1}
\end{equation}
Now any operator with $\hat S^2$ eigenvalue of $s(s+1)$ is now within the null space of $\mathcal{\hat O_s}$.

If we select $s=S+\frac12$ then
operate $\mathcal{\hat O}_{s=S+\frac{1}{2}}$ on both sides of (\ref{eq:sp1}), we get
\begin{align*}
    X(N,S+1/2,M_{S}+1/2) &= 1/c_{2} (S+M_{S}+1)\Theta[S,M_S]\otimes\ket{\alpha} \\
        &+\sqrt{(S+M_{S}+1)(S-M_{S})}X(N,S,M_{S}+1)\otimes\ket{\beta}
\end{align*}
Normalizing eliminates $c_{1}$.
\begin{align}
         &X(N,S+1/2,M_{S}+1/2) \\ =&\frac{1}{\sqrt{2S+1}}\left\{\sqrt{(S+M_{S}+1)}\Theta[S,M_S]\otimes\ket{\alpha}
        +\sqrt{(S-M_{S})}X(N,S,M_{S}+1)\otimes\ket{\beta}\right\}
        \nonumber
\end{align}
Similarly, operating $\mathcal{O}_{S-\frac{1}{2}}$ on \eqref{eq:sp1}
\begin{equation}
        X(N,S-1/2,M_{S}+1/2) = \frac{-\sqrt{(S-M_{S})}\Theta[S,M_S]\otimes\ket{\alpha}
        +\sqrt{(S+M_{S}+1)}X(N,S,M_{S}+1)\otimes\ket{\beta}}{\sqrt{2S+1}}
\end{equation}
Following the same strategy but adding $\ket{\beta}$ to $N$ electrons and applying
$\mathcal{O}_{S+\frac{1}{2}}$ and $\mathcal{O}_{S-\frac{1}{2}}$ on the product function

\begin{equation}
        X(N,S+1/2,M_{S}-1/2) = \frac{\sqrt{(S-M_{S}+1)}\Theta[S,M_S]\otimes\ket{\beta}
        +\sqrt{(S+M_{S})}X(N,S,M_{S}-1)\otimes\ket{\alpha}}{\sqrt{2S+1}}
\end{equation}

\begin{equation}
        X(N,S-1/2,M_{S}-1/2) = \frac{-\sqrt{(S+M_{S})}\Theta[S,M_S]\otimes\ket{\beta}
        +\sqrt{(S-M_{S}+1)}X(N,S,M_{S}-1)\otimes\ket{\alpha}}{\sqrt{2S+1}}
\end{equation}
This approach gives a genealogical way of constructing the spin-eigenfunction, which can be visualized using branching diagram (Fig. \ref{fig:branch}) and hence they are also known as branching diagram eigenfunctions. 

The x-axis of the branching diagram represents the number of electrons, and the y-axis represent the spin of the
system. Adding an electron is depicted by moving one unit in  +x-axis. Now the
added electron can increase the total spin by half a unit or decrease the spin by half. 
Addition spin is depicted by moving up 1/2 units in y-axis and subtraction
of spin is depicted by moving down 1/2 units down in y-axis. For example $N=5$ and
$S=1/2$, there can be five paths to reach the destination according to the
defined set of rules.  The possible paths are shown on the right of Fig. \ref{fig:branch}. Each one gives 
spin-eigenfunction, and they are orthogonal to each other. The left panel of
Fig. \ref{fig:branch} shows the number of pathways to reach $(5,\frac{1}{2})$.

\subsection{Example for $N=2 \rightarrow N=3$ }
The spin eigenfunctions are 
\begin{align*}
    X(2,0,0) &= \frac{1}{\sqrt{2}} [\alpha(1)\beta(2)-\beta(1)\alpha(2)] \\
    X(2,1,1) &= \alpha(1)\alpha(2)\\
    X(2,1,0) &= \frac{1}{\sqrt{2}} [\alpha(1)\beta(2)+\beta(1)\alpha(2)] \\
    X(2,1,-1) &= \beta(1)\beta(2)
\end{align*}
Incorporation of another electron to a two electron system coupled with $S=1$ has two possible values: $\dfrac{1}{2}$ or $\dfrac{3}{2}$. If $\alpha$ spin is introduced to the system, it leads to following set of eigenfunctions: 
\begin{align*}
    X(3,3/2,3/2) &=\alpha(1)\alpha(2)\alpha(3)\\
    X(3,3/2,1/2) &= \frac{1}{\sqrt{3}} [\alpha(1)\alpha(2)\beta(3) + \alpha(1)\beta(2)\alpha(3)+\beta(1)\alpha(2)\alpha(3)]\\
    X(3,3/2,-1/2) &=\frac{1}{\sqrt{3}} [\beta(1)\beta(2)\alpha(3) + \beta(1)\alpha(2)\beta(3)+\alpha(1)\beta(2)\beta(3)]\\
    X(3,3/2,-3/3) &=\beta(1)\beta(2)\beta(3)
\end{align*}
But there exist degeneracy for $S=1/2$, because $X(3,1/2,1/2)$ can be constructed from $X(2,1,1)$ by subtracting electron spin and from $X(2,0,0)$ by adding electron spin. 

\begin{align*}
    X(3,1/2,1/2;1) &= \frac{1}{\sqrt{6}}[2\times \alpha(1)\alpha(2)\beta(3) - \alpha(1)\beta(2)\alpha(3)-\beta(1)\alpha(2)\alpha(3)]\\
    X(3,1/2,1/2;2) &=\frac{1}{\sqrt{2}}[\alpha(1)\beta(2)\alpha(3)-\beta(1)\alpha(2)\alpha(3)]
\end{align*}

\section{Wigner projection operators}

Building on the orthogonality relation derived in the previous section, we can construct orthogonal projectors into an irreducible subspace.

We define the operators $\mathbf{P}^{[\lambda] IJ}$ as
\begin{equation}\label{eq:proj1}
\mathbf{P}^{[\lambda] IJ}= 
\sum_{g\in G} U^{[\lambda]}_{IJ}({g^{-1}}) \;{g}
\end{equation}
Since this operator is in the group algebra, we will need to consider it in a defining representation, say matrices $U(g)$. We will rely on the orthogonality theorem from the previous section to give the projectors into the irreducible subspace. An additional complication is introduced because we are not necessarily in an orthogonal basis.  

\subsection{Orthogonal basis}

To give an idea of our approach we begin with the orthogonal case first i.e. when the defining basis is $\{\ket{F_k}:\langle F_j\ket{F_i}=\delta_{ij}\}$. 
We will decompose the action
of the $g$ operator rightmost 
in \eqref{eq:proj1} using the unknown 
irreducible decomposition.  The summation over the irreducible representations closes after invoking the orthogonality theorems. The orthogonality theorem also fixes the components of the irreducible representation.
The action of group element $g$ can be represented on a fix basis e.g. by matrix $U(g)$.  Then the representation $U(g)$ can be written in the unknown irreducible representation decomposition giving
\begin{eqnarray}
U(g)=\sum^{\hat G}_\mu U^{[\mu]}(g)=\sum^{\hat G}_\mu \sum_{AB}^{d_\mu} U^{[\mu]}_{AB}(g)\ket{F^{[\mu]}_A}\bra{F^{[\mu]}_B}
\end{eqnarray}
After a little massaging, we will insert this expression into \eqref{eq:proj1}
\begin{eqnarray}
\mathbf{P}^{[\lambda] IJ}=
\sum_{g\in G}
U^{[\lambda]}_{IJ}({g^{-1}}) \;
\left(
\sum^{\hat G}_\mu 
\sum_{AB}^{d_\mu} U^{[\mu]}_{AB}(g)
\ket{F^{[\mu]}_A}\bra{F^{[\mu]}_B}
\right)
\end{eqnarray}
Applying the orthogonality relation: 
$\sum_g U^{[\mu]}_{im}(g^{-1})  U^{[\lambda]}_{nj}(g)
=\delta_{\lambda\mu}\delta_{ij}\delta_{mn}\;h/D_\lambda$ gives
\begin{eqnarray}
\mathbf{P}^{[\lambda]IJ}&=&
\sum^{\hat G}_\mu 
\sum_{AB}^{D_\mu}
\left(
\sum_{g\in G} U^{[\lambda]}_{IJ}({g^{-1}}) \;
U^{[\mu]}_{AB}(g)
\right)
\ket{F^{[\mu]}_A}\bra{F^{[\mu]}_B}\\
&=&
\sum^{\hat G}_\mu \sum_{AB}^{D_\lambda}
\left(\delta_{I B}\delta_{J A}\delta_{\lambda\mu} \frac h{D_\lambda}\right)
\ket{F^{[\mu]}_A}\bra{F^{[\mu]}_B}\\
&=&\frac h{D_\lambda}\ket{F^{[\lambda]}_J}\bra{F^{[\lambda]}_I }
\end{eqnarray}
Thus, we see that the matrix elements of operators in the irreducible representation are determined by the Wigner operators.

\end{document}